\documentclass[%
 aip,
 jcp,%
 amsmath,amssymb,floatfix,
 reprint,%
]{revtex4-1}

\usepackage[version=3]{mhchem} 

\usepackage{dcolumn}
\usepackage{bm}
\usepackage{todonotes}
\usepackage{tikz}
\usepackage{nicefrac}
\usepackage{array}
\usepackage{amssymb}
\usepackage{amsmath}
\usepackage{graphicx}
\usepackage{multirow}
\usepackage{color}
\usepackage{comment}
\usepackage[normalem]{ulem}
\usepackage{url}
\usepackage{mathrsfs}
\usepackage{subfigure}
\usepackage{float}

\newcommand*{\citen}[1]{%
  \begingroup
    \romannumeral-`\x 
    \setcitestyle{numbers}%
    \cite{#1}%
  \endgroup   
}

\newcommand{\tm}{\text{m}} \newcommand{\tf}{\text{f}}
\newcommand{\ts}{\text{s}}
\newcommand{\bq}{\mathbf{q}}

\begin{document}

\title{Decisive role of nuclear quantum effects on surface mediated water dissociation at finite temperature}

\author{Yair Litman}
\affiliation{Fritz Haber Institute of the Max Planck Society, Faradayweg 4--6, 14195 Berlin, Germany}

\author{Davide Donadio}
\affiliation{Department of Chemistry, University of California Davis, One Shields Ave. Davis, CA, 95616, USA}%
\affiliation{IKERBASQUE, Basque Foundation for Science, E-48011 Bilbao, Spain}
 
\author{Michele Ceriotti}
\affiliation{Laboratory of Computational Science and Modelling, \'Ecole Polytechnique F\'ed\'erale de Lausanne, Switzerland}
\author{Mariana Rossi}
\affiliation{Fritz Haber Institute of the Max Planck Society, Faradayweg 4--6, 14195 Berlin, Germany}

\date{\today}

\begin{abstract}
Water molecules adsorbed on inorganic substrates play an important role in several technological applications. 
In the presence of light atoms in adsorbates, nuclear quantum effects (NQE) influence the structural stability and the dynamical properties of these systems.
In this work, we explore the impact of NQE on the dissociation of water wires on stepped Pt(221) surfaces. 
By performing {\sl ab initio} molecular dynamics simulations with van der Waals corrected density functional theory, we note that several competing minima for both intact and dissociated structures are accessible at finite temperatures, making it important to assess whether harmonic estimates of the quantum free energy are sufficient to determine the relative stability of the different states. 
We thus perform {\sl ab initio} path integral molecular dynamics (PIMD) in order to calculate these contributions taking into account conformational entropy and
anharmonicities at finite temperatures. We propose that when when adsorption is weak and NQE on the substrate are negligible, PIMD simulations can be performed through a simple partition of the system, resulting in considerable computational savings.
We then calculate the full contribution of NQE to the free energies, including also anharmonic terms. We find that they result in an increase of up to 20\% of the quantum contribution to the dissociation free energy compared to the harmonic estimates.
We also find that the dissociation process has a negligible contribution from tunneling, but is dominated by ZPE, 
which can enhance the rate of dissociation by three orders of magnitude.
Finally we highlight how both temperature and NQE indirectly impact dipoles and the redistribution of electron density, causing work function to changes of up to 0.4 eV with respect to static estimates. This quantitative determination of the change in work function provides a possible approach to determine experimentally the most stable configurations of water oligomers on the stepped surfaces.
\end{abstract}

\maketitle

\section{Introduction \label{sec:intro}}

The characterization of the interface between water and metallic surfaces is an area that has been extensively explored theoretically and experimentally in the past decades. Such interest does not only stem from the catalytic processes in which these systems are involved, such as water gas shift reactions, natural gas steam reforming, and photocatalytic water splitting, but also from their ability to provide fundamental understanding on the structure of water at electrochemical interfaces, as well as on corrosion and wetting processes\cite{Thiel_1987_SurSciRep,Henderson_2002_SurSciRep,Verdaguer_2006_ChemRev}.
Despite the amount of studies and data available in the literature, as experimental setups and theoretical methods advance, new structures of water at these interfaces are still found and our understanding of which fundamental physical aspects stabilize certain structures and processes is revised\cite{Hodgson_2009_SurSciRep,Carrasco_NatMat_2012}.

For this class of systems, especially when water dissociation is involved, a potential energy surface 
based on electronic structure theory can provide 
quantitative physical understanding. 
Theoretical studies of water adsorbed on metallic surfaces, however, are most often based on static calculations of the potential energy, without accounting for temperature effects. 
This approach, albeit successful at times\cite{Donadio_JACS_2012,Carrasco_NatMat_2012}, encounters limitations when barriers between different 
structural motifs are shallow, allowing the system to adopt many different conformations at a given temperature, which are often experimentally relevant.
Anharmonic connections among different minima in the potential energy surface call for approaches beyond the harmonic approximation.
Moreover, the structural, dynamic, and electronic properties of water are known to be heavily affected by the quantum nature of the nuclei even at room temperature\cite{gibe+14jpcb,Ceriotti_ChemRev_2016,Ceriotti_PNAS_2013,chen+16prl,Wilkins_JPCL_2017,rybk-vand17jpcl,RossiCeriotti2014,RossiMano2016}. In fact, nuclear quantum effects (NQE) have also been shown, through several experiments and a few theoretical works, to play a crucial role in the behaviour of organic adsorbates on metallic surfaces\cite{Zhang_SurSci_2011,Lauhon_PhysRevLett_2000,McIntosh_PhysChemLet_2013, Kyriakou_AcsNano_2014,Kumagai_PhysRevB_2009,Mitsui_Science_2002,Ranea_PhysRevLett_2004,Kumagai_PhysRevLett_2008,Koitaya_ChemRec_2014,Meng_Nature_2015}. 
It is thus to be expected that both conformational entropy and nuclear quantum contributions impact the physics underlying the processes of water adsorption and dissociation on metallic surfaces.

In this paper we address the dissociation of water at Pt(221) surfaces as a model system of interest. This surface is relevant due to its stepped geometry \cite{Woodruff_2016_CondMatt}, since it is well established that surface defects like steps or vacancies are reactive centers where most surface reactions happen. It has also been previously shown, based on minimum energy calculations where harmonic zero point energies (ZPE) were added, that ZPE can be decisive in favoring dissociated water structures in this system \cite{Donadio_JACS_2012}. We here present a study, based on {\sl ab initio} molecular dynamics (MD) and {\sl ab initio} path integral molecular dynamics (PIMD) where we establish the magnitude of the NQE contributions to various aspects of the dissociation process at finite temperature and discuss the impact of these effects on electronic structure properties of the system, like charge rearrangements and work function changes. Due to the high cost especially of {\sl ab initio} PIMD simulations, we also propose a simple and general scheme based on a spatial partition of the system, which is combined with ring-polymer contraction~\cite{mark-mano08jcp,mark-mano08cpl} to accelerate PIMD simulations when only physisorption of the adsorbate is involved and one is interested in NQE in the adsorbate. 

This paper is organized as follows: In section \ref{sec:methods}, we
present the details of the simulations and define our spatially localized contraction scheme (SLC).
In sections \ref{sec:results:stat} to \ref{sec:pathway} we present and discuss our results for static calculations of different minima and the characterization of dissociation pathways and rates. 
We present MD and PIMD results in \ref{sec:results:therm} where we discuss the quantum contributions to the free energy of dissociation, and in section \ref{sec:results:dyn} we present finite-temperature 
geometrical and electronic properties of the system.  
Finally, in section \ref{sec:conclusions} we summarize the most important conclusions of the present work.

\section{Methods \label{sec:methods}}

\subsection{Static evaluations of structures and energies \label{sec:methods:stat}}

For all the static structure and energy calculations 
we have modeled our system of water
adsorbed on Pt(221) using a unit cell containing one step and two water molecules, with periodic boundary conditions. We have aligned the surface perpendicular to the $z$ direction. We have imposed a vacuum of 60 \AA\, and a dipole correction \cite{Neugebauer_1992_PRB} in order to isolate the systems in the $z$ direction. Unless otherwise specified, our calculations were performed using a 4 layer-thick slab, where only the first two top layers were allowed to relax and the others were kept fixed in their bulk positions, as well as a 4x4x1 k-point sampling. We compare our results to a converged simulation of 8 surface layers and 20x20x1 k-points in order to determine the magnitude of our remaining errors.

Energies, electronic densities, and forces were calculated through density-functional theory (DFT) in the generalized gradient approximation, using the functional by Perdew, Burke, and Ernzerhof (PBE)\cite{PBE} with a pairwise van der Waals correction (vdW) specifically tailored for surface calculations \cite{Ruiz_PRL_2012}. In our calculations we only include pairwise vdW corrections within the adsorbate and between the adsorbate and the surface, thus ignoring the dispersion forces among Pt atoms. This proved necessary due to the lack of electronic screening in this correction, which would produce unphysically short lattice constants for Pt. 
All the electronic structure calculations are performed using the all-electron FHI-aims program \cite{FHI-AIMS} with both {\it light} and {\it tight} basis sets and numerical settings. 
.

Reaction paths were found with the string method\cite{Weinan_JCP_2007} combined with
the climbing image technique\cite{Graeme_JCP_2000} using 9 replicas. 
Charge distribution analysis was performed with the Hirshfeld partition scheme\cite{Hirshfeld_1977_TheoChimActa}, unless otherwise specified.

\subsection{Ensemble average simulations \label{sec:methods:ensemble}}

To calculate ensemble average properties with both classical and quantum nuclei we have used the i-PI code \cite{i-pi}.
The forces to evolve molecular dynamics trajectories were obtained from the FHI-aims code and passed to i-PI through an interface
based on internet sockets. For dynamics, we have modelled our system with {\it light} numerical and basis sets settings  \cite{FHI-AIMS} 
and used four layers for the slabs. For the {\sl ab initio} molecular dynamics simulations, we coupled our system to a Langevin thermostat ($\tau$ = 40 fs) and 
for the {\sl ab initio} path integral molecular dynamics simulations we coupled the ring polymer to the colored noise PIGLET thermostat \cite{Ceriotti_PRL_2012},
thus obtaining converged results using 6 and 12 beads for 300~K and 160K simulations respectively.
In all cases we used a 0.5 fs time step for the integration of the equations of motion.
The simulations were run for at least 10 ps and at least 5 ps for 300~K and 160~K, respectively.

\subsection{Spatially Localized Contraction \label{sec:methods:slc}}

\begin{figure*}
\centering
\includegraphics[width=0.8\textwidth]{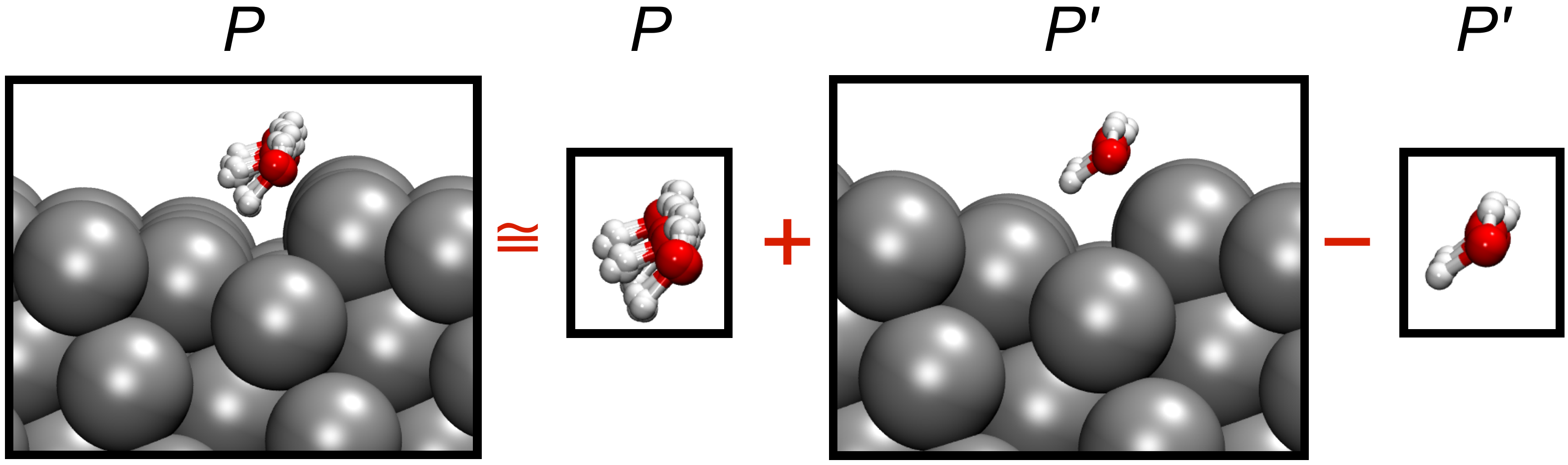}
\caption{Schematic representation of the spatially localized contraction. $P$ and $P'$ represent the number of beads used to calculate each system. \label{fig:contrac-scheme}}
\end{figure*}

The purpose of this section is to present a simple scheme that can be used to accelerate path integral molecular dynamics simulations of weakly bound surface adsorbates. Ideas involving spatial partitions of systems in order to treat different regions with different methods have been previously proposed in the literature.\cite{poma-dell10prl} For example, in the approach by Kreis \textit{et. al} \cite{Kreis_JCTC_2016} a position dependent mass Hamiltonian is used, which allows the definition of a quantum, a hybrid and a classical region. However, in surface science applications,  where a ``natural" partition of space is possible, the essential idea can be implemented much more transparently, using a ring-polymer contraction scheme in which the adsorbate is treated fully quantum mechanically in the gas phase, and the interaction with the surface (that is often mediated by weak non-covalent forces) is treated with a reduced number of beads, as we show below.

In the following, we consider the ring polymer Hamiltonian, given by
\begin{multline}
   \label{eq:H_P}
   H_P ( \{ p^{(k)}_j\}, \{q^{(k)}_j\}) =\\
  \sum_{j=1}^{3N} \sum_{k=1}^P \left[ \frac {( p_j^{(k)})^2 }{2m_j}
   + \frac{m_j\omega_P^2}{2} ({q}_j^{(k)}-{q}_j^{(k+1)})^2 \right] +  \\ 
   \sum_{k=1}^P V_\tf^{(k)}( q_1^{(k)},  q_2^{(k)}, \dots,  q_{3N}^{(k)}),
\end{multline}
where $q_j^{(k)}$ and  $p_j^{(k)}$
represent the position and momentum of the $j$-th degree of freedom (representing both particle index and Cartesian direction),
at the imaginary time slice $k$, $({\bq}^{(P+1)}\equiv {\bq}^{(1)}$), and $\omega_P = (\beta_P\hbar)^{-1}$ with $(k_BT)^{-1} = P \beta_P$.
The full potential $V_\tf$ depends on both the positions of atoms in the molecule $\bq_\tm$ and those of the surface, $\bq_\ts$. Usually the surface constitutes the largest fraction of the system, and it involves heavy atoms that are weakly quantized. 
In these cases it can be useful to apply a ring-polymer contraction (RPC) scheme, in which the full potential is computed on a reduced number of replicas $P'$, and the fully-quantized calculation is performed only on the isolated molecule. In other terms, the full ring-polymer potential $V_P(\bq) =\sum_{k=1}^P V_\tf(\bq^{(k)})$ is approximated as a ``spatially-localized contraction'' (SLC) potential,
\begin{equation}
\tilde{V}_P(\bq) \approx 
\frac{P}{P'}\sum_{k=1}^{P'} \left[V_\tf(\tilde{\bq}^{(k)})-V_\tm(\tilde{\bq}_\tm^{(k)})\right]
+ \sum_{k=1}^P V_\tm(\bq_\tm^{(k)}),
\label{eq:v-slc}
\end{equation}
where $\tilde{\bq}_\tm^{(k)}$ refers to the coordinates of a ``contracted'' ring polymer obtained by Fourier interpolation of the full ring polymer \cite{mark-mano08jcp}. A schematic representation of our scheme can be found in Fig. \ref{fig:contrac-scheme}. We have included this partition scheme in the i-PI code, combining it seamlessly with the RPC/Multiple-time-stepping (MTS) implementation~\cite{kapi+16jcp}.

An error analysis for this procedure -- that applies more in general to any Fourier RPC scheme -- is discussed in Appendix~\ref{app:rpc-harmonic}. To provide a more physical understanding of the approximations involved in the use of Eq.~\eqref{eq:v-slc},
let us assume that one can formally break down the full potential ($V_\tf$) into two fragments corresponding to the molecule and the surface ($V_\tm, V_\ts$) and an interfragment component ($V_{\tm\ts}$):

\begin{equation}
V_\tf(\bq) = V_\ts(\bq_\ts) + V_\tm(\bq_\tm) + V_{\tm\ts}(\bq_\ts,\bq_\tm).
\end{equation}
The SLC does not require this decomposition to be performed explicitly since one only computes the full potential and the energy of the isolated molecular fragment, but this is a useful illustrative analysis.

Adding and subtracting the potential evaluated on the contracted ring polymer from that computed on $P$ beads one can separate the different contributions:
\begin{eqnarray}
\label{eq:full_H}
V_P(\bq) =\sum_{k=1}^P V_\tf(\bq^{(k)}) + (1-1)
\frac{P}{P'}\sum_{k=1}^{P'} V_\tf(\tilde{\bq}^{(k)}) = \\
\frac{P}{P'}\sum_{k=1}^{P'} V_\tf(\tilde{\bq}^{(k)}) + \nonumber \\
\label{eq:full_H2}
\sum_{k=1}^P V_\tm(\tilde{\bq}_\tm^{(k)}) -
\frac{P}{P'}\sum_{k=1}^{P'} V_\tm(\tilde{\bq}_\tm^{(k)}) +\\
\label{eq:full_Vs}
\sum_{k=1}^P V_\ts(\tilde{\bq}_\ts^{(k)}) -
\frac{P}{P'}\sum_{k=1}^{P'} V_\ts(\tilde{\bq}_\ts^{(k)}) +\\
\label{eq:full_Vms}
\sum_{k=1}^P V_{\tm\ts}(\tilde{\bq}_\ts^{(k)},\tilde{\bq}_\tm^{(k)}) -
\frac{P}{P'}\sum_{k=1}^{P'} V_{\tm\ts}(\tilde{\bq}_\ts^{(k)},\tilde{\bq}_\tm^{(k)}).
\end{eqnarray}
One can see that taking the SLC approximation amounts to neglecting the terms in Eqs.~\eqref{eq:full_Vs} and~\eqref{eq:full_Vms}. 
Neglecting the quantization of the surface is very well justified whenever the atoms are heavy and behave classically (e.g. when studying adsorption on a transition metal, as we do here) or when the quantization of the surface does not change its reactivity. In the latter case, the quantization of the surface would impact the absolute quantum thermodynamics of the system, but would not be important if one focuses on the adsorbate or the adsorption process. An example of a case in which Eq.~\eqref{eq:full_Vms} cannot be neglected would be that of a hydroxilated surface that changes its pKa due to quantum fluctuations.
Neglecting the interfragment term~\eqref{eq:full_Vms} can only be justified when the interaction involves  relatively weak interactions that do not change dramatically the high-frequency vibrations  of the molecule. As we will discuss below, here we present examples for which this approximation is justified, and some for which it breaks down, providing a practical demonstration of the scope of applicability of the SLC technique.

\begin{figure*}[ht]
    \includegraphics[width=0.9\textwidth]{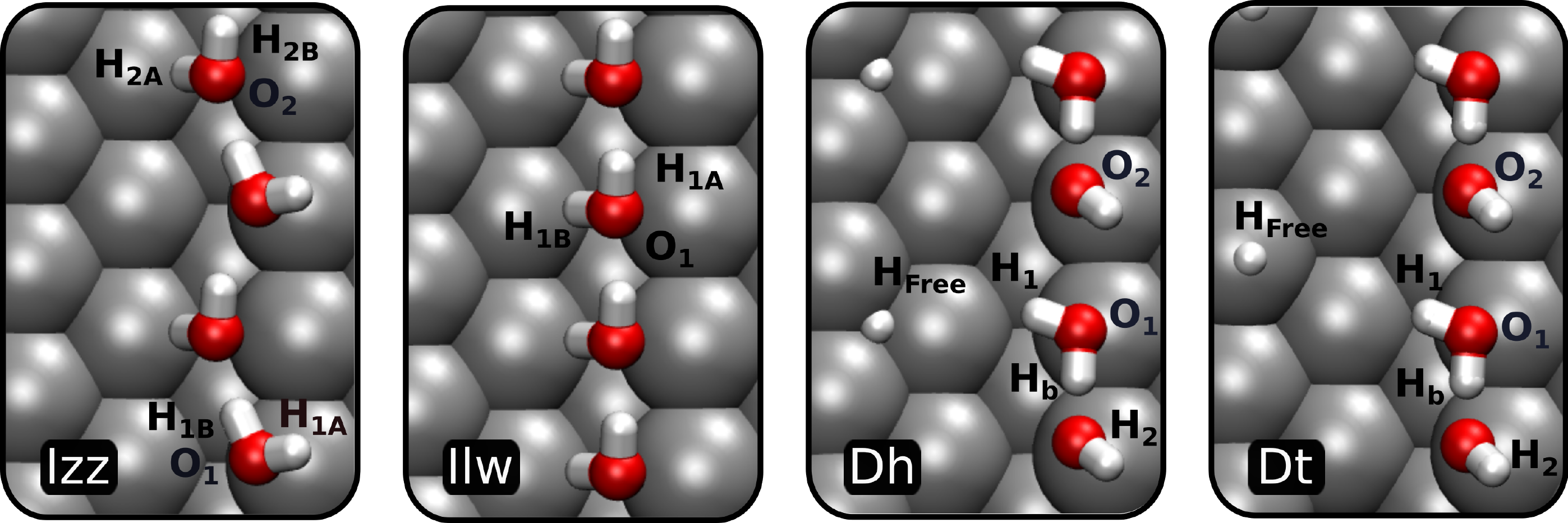}
    \caption{Geometries of the local minima discussed in this work, from left to right:  The intact ``zigzag" structure Izz,  the intact ``L-wire" structure Ilw,  the dissociated structure with the hydrogen on the hollow site Dh, and  the dissociated structure with the hydrogen on the top site Dt. } \label{fig:structures}
\end{figure*}

\section{Results and Discussion \label{Results}}

\subsection{Minimum energy structures \label{sec:results:stat}}

We performed exploratory molecular dynamics runs at 300~K 
starting from intact and partially dissociated configurations of the water dimer.
During our MD simulations the system visited several different, seemingly meta-stable, conformations.
In particular, beyond the structures already reported in the literature for this system \cite{Donadio_JACS_2012}, we found two relevant different structures, one for the dissociated case and the other
for the intact case. We show all structures in Fig. \ref{fig:structures}, and coin them
``Izz" and ``Ilw" for intact water molecules, and  Dt and Dh for the
dissociated water molecules. These labels were chosen given the shape of the pattern formed by
the water molecules on the step and the preferred adsorption site of the dissociated hydrogen, namely a zigzag-wire and an ``L"-wire
for the intact cases, and the preference of the dissociated hydrogen to adsorb on the top or fcc-hollow site for the dissociated structures.
We note that the Ilw structure is also reported in Ref.~\cite{Kolb_PCCP_2016} on the Pt(533) surface. For these four relevant structures we performed geometry minimizations and computed ZPE and vibrational free energies in the (quantum) harmonic approximation at 0 K and two other pertinent temperatures: 160K, the experimental temperature where the water chain can be isolated by thermal programmed desorption (TPD) \cite{Morgenstern_PRL_1996,Nakamura_JPCC_2009,Grecea_JPCB_2004}, and 300~K -- room temperature. The results are shown in Table \ref{tab:Minimum_all}. Our comparison between the results obtained with {\it light} settings, {\it tight} settings  for basis sets and numerical grids, and a converged model containing 8 surface layers and a $20\times20\times1$ k-point mesh shows that there is a fortuitous error cancellation in the numbers obtained with {\it light} settings, and thus at this level our errors are of about 30 meV on relative energies. For the subsequent dynamics discussed in Section \ref{sec:results:dyn} we cannot afford {\it tight} settings or a larger slab. We also note that on the ``converged" level, adding van der Waals contributions to this system strengthens the H-bonds  and makes the intact Izz structure 15 meV more stable with respect to the dissociated one. In other studies of water adsorbed on flat Pt surfaces and on Pt(211), this effect has been reported to be negligible\cite{Carrasco_2013_JCP,Pekoz_JPCC_2017}.

Regarding the intact-molecules case, the Ilw structure is computed to be more stable than the Izz structure at any temperature within our model by up to around 100 meV per water dimer. Experimental data  from oxygen K-NEXAFS spectra suggest a chain on the step composed of one hydrogen donation and one hydrogen acceptance per molecule with two different hydrogen bond lengths \cite{Endo_JPCC_2012}, thus pointing to the Izz structure. Also, based on X-Ray experiments\cite{Nakamura_JPCC_2009}, the Izz structure has been suggested due to the measured positions of the O atoms. In experiments, these wires are obtained through evaporation of water molecules from a surface that was fully covered, but it is also known from the literature that the addition of one water molecule on the terrace makes the Ilw structure convert to the Izz structure \cite{Kolb_PCCP_2016}. We propose that the possible absence of Ilw structures in experimental measurements could be due to a kinetic trapping of Izz structures or to the presence of only dissociated structures (we have checked that when considering larger supercells the Ilw structure is still the most stable).

When the molecules are dissociated, we find that the dissociated hydrogen atom prefers to adsorb on top sites (Dt) at any temperature. Even though within our simulation setup (functional, basis-sets, etc.) we find isolated hydrogens to prefer the fcc-hollow site on clean Pt(111) surfaces, in agreement with the literature\cite{Watson_JPB_2001,Phatak_JPC_2009,Koeleman_PRL_1986}, we have thoroughly checked that this is not the case for the Pt(221) surface studied here.

Our simulations prompt two further observations regarding the structural patterns of this system. The first is that we found cases where, for the wire containing intact water molecules,
one of the molecules occupied a position on the terrace, thus breaking the water wire periodicity (see SI). This ``dimerization" is also observed in simulations of other stepped surfaces\cite{Pekoz_JPCC_2017}. The other observation is that when considering a lower coverage of water on the surface, in which step sites can be empty, the dissociated hydrogen prefers to adsorb on the step over any terrace site, with an energy stabilization of around 0.1 eV.

\begin{table}[htbp]
    \caption{Potential and harmonic free energies (in eV) for the four local minima studied in this work (see text), calculated with the PBE+vdW$^{\text{surf}}$ exchange correlation functional, {\it light} and {\it tight} settings for basis sets and numerical settings, and $4\times4\times1$ k-points. The settings corresponding to ``converged" were calculated with 8 layers for the slab, {\it tight} settings and $20\times20\times$1 k-points. We set to zero the potential energy of the Izz structure at the {\it light}, {\it tight}, and ``converged" settings. \label{tab:Minimum_all}}
      \centering
        \begin{tabular}{c|c|cccc}
            \hline
                                     &         &   Izz     & Ilw                 & Dh                  & Dt                     \\
            \hline 
            \hline
            \multirow{4}{*}{{\it light} settings }&$V$  & \multicolumn{1}{l}{0.00} & \multicolumn{1}{l}{-0.10}&  \multicolumn{1}{l}{0.18}  & \multicolumn{1}{l}{0.05}\\
                                 & $F^{Ha}_{0K}$ & \multicolumn{1}{l}{1.95} & \multicolumn{1}{l}{1.85}&   \multicolumn{1}{l}{1.94}  & \multicolumn{1}{l}{1.86}\\
                                 &$F^{Ha}_{160K}$& \multicolumn{1}{l}{1.16} & \multicolumn{1}{l}{1.07}&   \multicolumn{1}{l}{1.17}  & \multicolumn{1}{l}{1.09}\\
                                 &$F^{Ha}_{300K}$& \multicolumn{1}{l}{-0.67} & \multicolumn{1}{l}{-0.74}& \multicolumn{1}{l}{-0.63} & \multicolumn{1}{l}{-0.71} \\
            \hline
            \hline
            \multirow{4}{*}{{\it tight} settings} &$V$  & \multicolumn{1}{l}{0.00} & \multicolumn{1}{l}{-0.03}&  \multicolumn{1}{l}{0.12}  & \multicolumn{1}{l}{0.01}\\
                                 & $F^{Ha}_{0K}$ & \multicolumn{1}{l}{1.93} & \multicolumn{1}{l}{1.91}&   \multicolumn{1}{l}{1.88}  & \multicolumn{1}{l}{1.82}\\
                                 &$F^{Ha}_{160K}$& \multicolumn{1}{l}{1.13} & \multicolumn{1}{l}{1.10}&   \multicolumn{1}{l}{1.08}  & \multicolumn{1}{l}{1.03}\\
                                 &$F^{Ha}_{300K}$& \multicolumn{1}{l}{-0.71} & \multicolumn{1}{l}{-0.75}& \multicolumn{1}{l}{-0.75} & \multicolumn{1}{l}{-0.79} \\
                        \hline
                        \hline
            Converged &$V$  & \multicolumn{1}{l}{0.00} & \multicolumn{1}{l}{-0.09}&  \multicolumn{1}{l}{0.13}  & \multicolumn{1}{l}{0.08}\\
            \hline           
        \end{tabular}
\end{table}

\begin{figure}[!hbt]
   \includegraphics[width=\columnwidth]{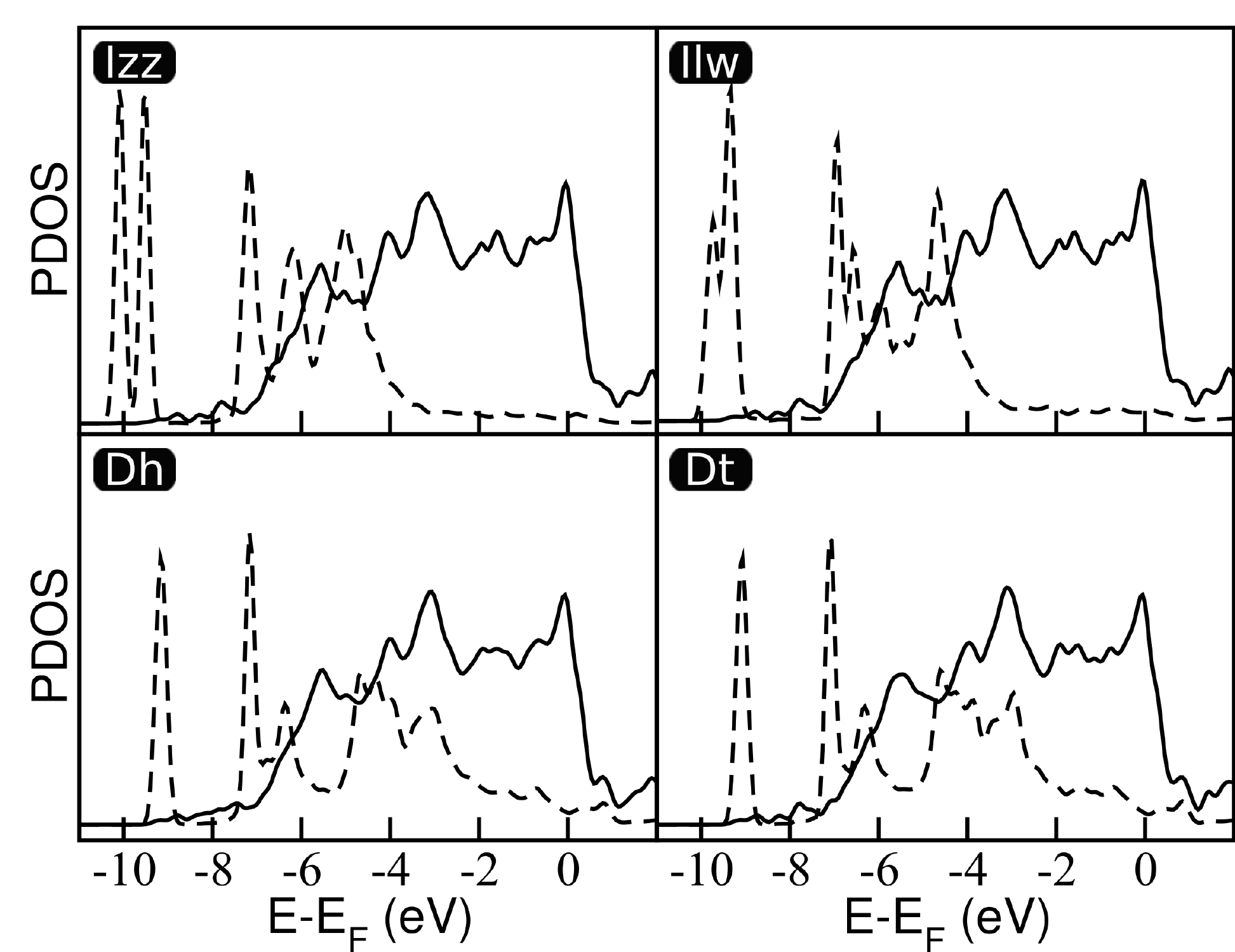}
   \caption{Projected density of states (PDOS) on the Pt states (solid line) and O states (dashed line). Different structures are labeled in the figure.
   }
   \label{fig:pdos-light}
\end{figure}

Finally, regarding the electronic structure of these systems, we show in Fig. \ref{fig:pdos-light} the projected density of states (PDOS) for the four local-minima discussed in Table \ref{tab:Minimum_all}. As reported previously,\cite{Pekoz_JPCC_2014} upon dissociation a strong orbital hybridization takes place involving the oxygen states, for both Dt and Dh geometries. Regarding the charge state of the adsorbates, we have performed Hirshfeld and Bader charge analysis\cite{TangHenkelman2009} (see SI), finding that both charge analysis show a slight net positive charge on the water molecules when they are intact, and when they are dissociated the hydrogen is completely screened (no charge) and there is a slight negative charge (-0.1 to -0.3 $e$) on the remaining H$_3$O$_2$ complex.

We conclude this section stressing that both at the experimental temperature of 160 K and at room temperature there are at least a few structures that are thermodynamically accessible with statistically significant populations, making it important to evaluate accurately the impact of anharmonicities and nuclear quantum effects.

\subsection{Vibrational Analysis of the Zero Point Energy Effect on Dissociation}

\begin{table}[htb]
        \caption{Dissociation energies in eV. Light settings, 4x4x1, 4-layers and the PBE+vdW$^{\text{surf}}$ functional.}
        \label{tab:Dis_light}
        \begin{center}
        \begin{tabular}{ccccc}
        \hline
        Structures &  E$_d$ & $\Delta F^{0K}_{d}$ & $\Delta F^{160K}_{d}$ & $\Delta F^{300K}_{d}$ \\
        \hline
        \hline
        Izz$\to$Dt     &  0.05 & -0.08  & -0.07  &-0.04  \\
        Izz$\to$Dh   &  0.18 & -0.01  & 0.01   & 0.04  \\
        Ilw$\to$Dt    &  0.15 &  0.02  & 0.02   & 0.03  \\
        Ilw$\to$Dh &  0.28 &  0.10  & 0.10   & 0.11  \\
        Izz$\to$Dh [\citen{Donadio_JACS_2012}] & 0.17  &-0.01   & -   & -  \\
        \hline
        \end{tabular}
        \end{center}
\end{table}

After characterizing the different minima, we show in Table  \ref{tab:Dis_light} the dissociation energies and dissociation free energies
($E_d = V_{\text{diss}} - V_{\text{intact}}$ and $\Delta F_d = F_{\text{diss}} - F_{\text{intact}}$) for the different geometries with {\it light} settings, 
and $4\times4\times1$ k-points since this is the level of theory we will use
for the dynamics in the next sections and, as discussed in the previous section, 
due to a seemingly fortuitous cancellation of errors, this level of theory gives close to converged results. 
The dissociation reaction is endoenergetic at the potential energy and only becomes exoenergetic with the addition of ZPE for the Izz$\to$Dt/Dh reactions. 
The Ilw$\to$Dt reaction is only mildly endoenergetic ($\approx 20$meV) when including ZPE and temperature effects (which is within our error bars). 
We observe that the  Izz$\to$Dt reaction is the most favorable at all temperatures. 
We note that even though only the Izz$\to$Dh reaction was previously reported in the literature, our numbers are in good agreement to the ones reported in Ref. \cite{Donadio_JACS_2012}.

\begin{figure}[!hbt]
    \includegraphics[width=\columnwidth]{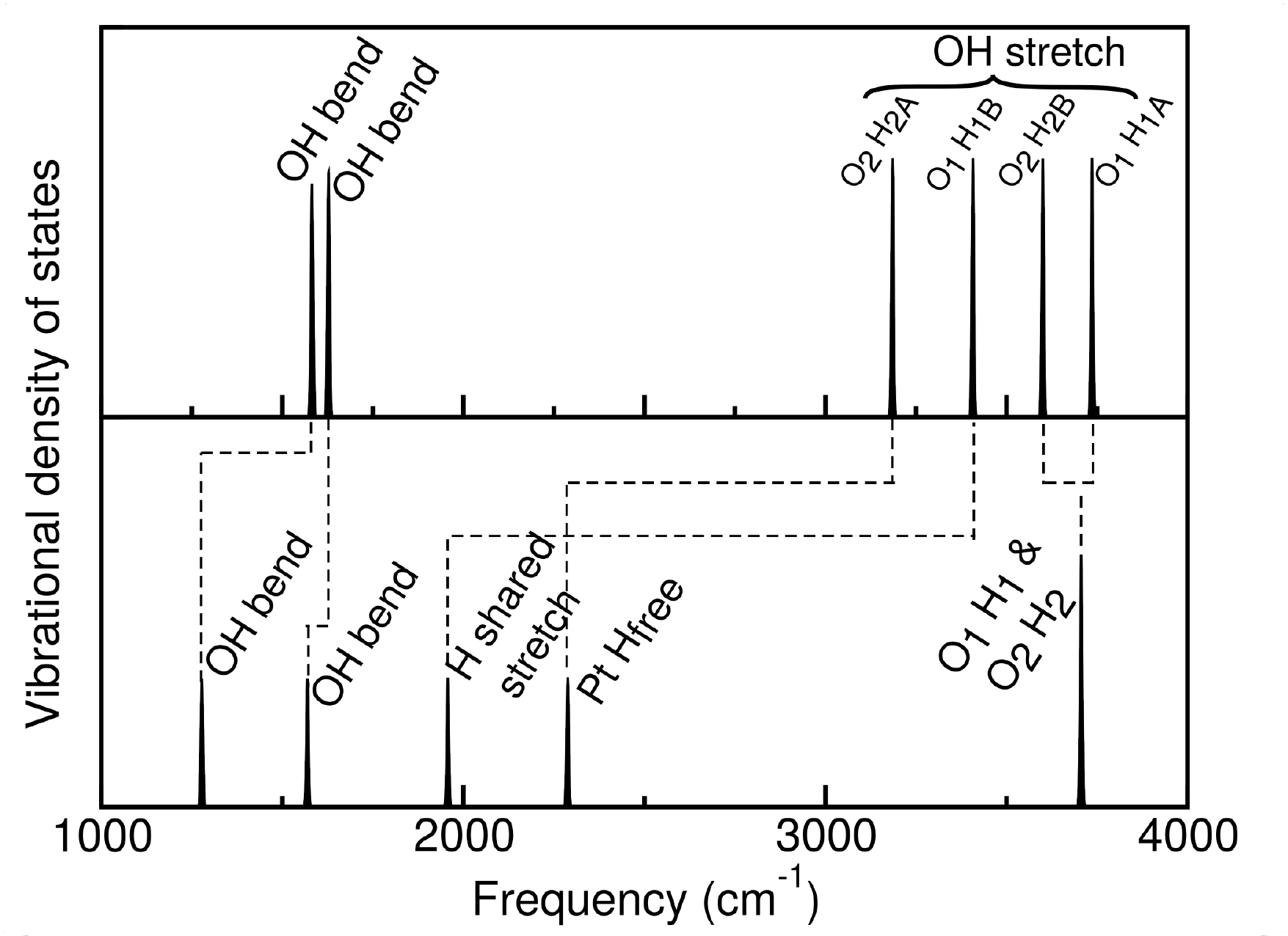}     
    \caption{Vibrational density of states of the Izz structure (top) and the Dt structure (bottom).} \label{fig:vdos}
\end{figure}

The role of ZPE can be understood in detail by analyzing the normal mode frequencies of intact and dissociated structures. The vibrational density of states for the Izz and Dt structures, shown in Fig. \ref{fig:vdos}, suggest that the ZPE stabilization of the dissociated structures can be  traced back exclusively to the water vibrational modes and the H-Pt stretch vibration.
In the intact structure, there are two bending modes  at about 1600~cm$^{-1}$, one OH stretching mode involved in the bond formation with the surface at about 3200 cm$^{-1}$, 
two OH stretching modes which are involved in H-bonds at about 3400 and 3600~cm$^{-1}$, and, finally, 
one OH stretching mode which corresponds to the to the hydrogen that doesn't participate in any bond (``free OH") at about 3700 cm$^{-1}$. 
Upon dissociation, the two highest stretching modes merge into a quasi-degenerate peak at about 3700cm$^{-1}$, which corresponds to the two  free OH stretching modes,
the stretching mode of the shorter H-bond that was around $3400$ cm$^{-1}$ converts into the oscillation of the shared H around 1950cm$^{-1}$, representing a massive decrease in this frequency and thus in the associated zero-point energy, 
and finally the stretching mode of the H atom that dissociates, which was at  about $3200$ cm$^{-1}$, turns into the Pt-H stretching at about 2300 cm$^{-1}$, representing a further decrease in the associated ZPE.
Bending modes remain similar but also present a slight decrease in frequencies as can be seen in Fig. \ref{fig:vdos}.
Thus the vibrational modes of water make up for a decrease in ZPE of 0.16 eV, which represents around 85\% of the total  ZPE difference (0.19 eV).
The other structures follow a similar pattern.

For collective modes of the adsorbate corresponding to their rotations and translations, which lie between 50 and 150 cm$^{-1}$, we observe that upon dissociation these become stiffer (blue-shifted) by around 10 to 40 cm$^{-1}$ if compared to the intact adsorbated molecules. The dissociated hydrogen also presents translational modes around 380 cm$^{-1}$. These modes have less of an impact on ZPE contributions, but can be important for entropic contributions with increasing temperature\cite{RossiBlum2013}, which should disfavor dissociation.

\subsection{Dissociation Pathways and Barriers \label{sec:pathway}}

We computed the minimum energy path of the Izz$\to$Dt and of the Ilw$\to$Dt reactions.
We obtained a barrier of 0.41 eV for the most favorable pathway of Izz$\to$Dt, which involves the dissociation of the hydrogen labeled H$_{2A}$ in Fig \ref{fig:structures}, and a barrier of 0.43 eV for the Ilw$\to$Dt reaction. 
Detaching H$_{1A}$ instead, produces a dissociation barrier that can be almost 1 eV higher. We note that the inclusion of vdW interactions actually lowers the dissociation barrier by 60 meV -- a truly non-negligible contribution.
With our calculated barriers and using transition state theory \cite{Eyring_1935_JCP,Miller1993}, we computed the reaction rate $k_d$ in the harmonic approximation  (hTST) 
and in the quantum quasi-harmonic approximation  (q-hTST) at 160 K. While hTST uses the classical harmonic partition function to describe reactants and products, q-hTST uses the quantum harmonic partition function. 
The expressions for the rate are the following,
\begin{eqnarray} \label{eq:hTST}
  k^{\text{hTST}} &=& \frac{1}{2\pi}
  \frac{\prod_{i=1}^{N}\omega_i^{\text{min}}}{\prod_{i=1}^{N-1}\omega_i^{\text{sp}}}
  e^{-\beta \Delta V} \\
  \label{eq:qhTST}
  k^{\text{q-hTST}} &=& \frac{1}{2\pi\hbar\beta}
  \frac{\prod_{i=1}^{N} 2\text{sinh}(\frac{\beta \hbar\omega_i^{\text{min}}}{2})}
       {\prod_{i=1}^{N-1} 2\text{sinh}(\frac{\beta \hbar\omega_i^{\text{sp}}}{2})}
  e^{-\beta \Delta V}
\end{eqnarray}
where $N$ is the number of vibrational degrees of freedom in the reactant, $\beta=1/k_BT$, $\omega_i^{\text{min}}$ and $\omega_i^{\text{sp}}$ are the phonon frequencies of the minimum and the saddle point respectively, and $\Delta V=V_{\text{sp}}-V_{\text{min}}$.

We obtain $k_d^{\text{hTST}}=0.2$ ($0.1$) s$^{-1}$ and $k_d^{\text{q-hTST}}=1240$ ($289$) s$^{-1}$ for the Izz$\to$Dt (Ilw$\to$Dt) reaction, pointing towards a major role of ZPE on the dissociation reaction of these systems. It should be stressed that q-hTST only takes (approximately) into consideration the quantum correction originated by the ZPE, neglecting tunnelling effects. This approximation turns out to be a good one in this case, since
we obtain a crossover temperature  (defined by $T_c=\hbar\omega_{sp}/2\pi k_b$ where $\omega_{sp}$ is the frequency of the first order saddle point) of $T_c=132$ ($124$) K, which is around 30~K lower than the typical experimental temperature, such that tunnelling effects are not expected to play a major role at the condition of most experiments. 

\subsection{Quantum contributions to the free energy of dissociation \label{sec:results:therm}}

We investigated the impact of NQE on the free energy of dissociation at finite temperatures. In order to calculate the contribution of NQE to the free energies, we used a classical to quantum thermodynamic integration given by\cite{Rossi_PCL_2015,FangMichaelides2016}
\begin{eqnarray}
\label{eq:TI}
  \Delta F_q = 
\int_{0}^{1} \frac{2\langle K(m_0/g^2)-K_{class}\rangle}{g} dg, 
  \end{eqnarray} 
where $K$ is the kinetic energy, $m_0$ is the physical mass of the
atoms, $\mu$ the mass integration variable, and $g = \sqrt{m_0 / \mu}$, a change of variables that makes the integrand better behaved \cite{Ceriotti_2013_JCP, Rossi_PCL_2015}. The
brackets denote an ensemble average, and the masses of all atoms are scaled in this procedure.

\begin{table}[htbp]
    \caption{Quantum kinetic energy $K$ of the full systems containing dissociated and intact water molecules in PIMD simulations at 300~K. Energy in eV.  \label{tab:SLC}}
    \begin{center}
    \begin{tabular}{ccc}
    \hline
       System &  SLC  &  $K$  \\
    \hline
    \hline
        Intact      &   None       &    1.963 $\pm$  0.003     \\
        Intact      & 1 bead       &    1.960 $\pm$ 0.002     \\
        Dissociated &   None       &   1.880 $\pm$ 0.007      \\
        Dissociated & 1 bead       &   1.742 $\pm$ 0.002      \\
        Dissociated & 4 beads      &   1.847 $\pm$ 0.002      \\
    \hline
    \end{tabular}
    \end{center}
\end{table}

In order to perform our PIMD simulations in the most efficient way possible, here we make use of the SLC scheme introduced in section \ref{sec:methods:slc}, after verifying that it gives sensible results for our systems. 
For this verification, we report in Table \ref{tab:SLC} the total quantum kinetic energy calculated with the centroid kinetic virial estimator 
for both dissociated and intact systems at different levels of spatial contraction on the full system, keeping the water molecules always evaluated with 6 beads
and the PIGLET thermostat  at 300~K. We compare our results with simulations where the whole system was calculated with 6 beads and the PIGLET thermostat. 
While in the intact case the contraction to the centroid for the full system results in essentially no difference in total quantum kinetic
energy, in the dissociated case, even when calculating the full system with 4 beads, one still does not recover the full quantum kinetic energy, making an error of more than 30 meV. 

\begin{figure}[!hbt] 
    \includegraphics[width=1\columnwidth]{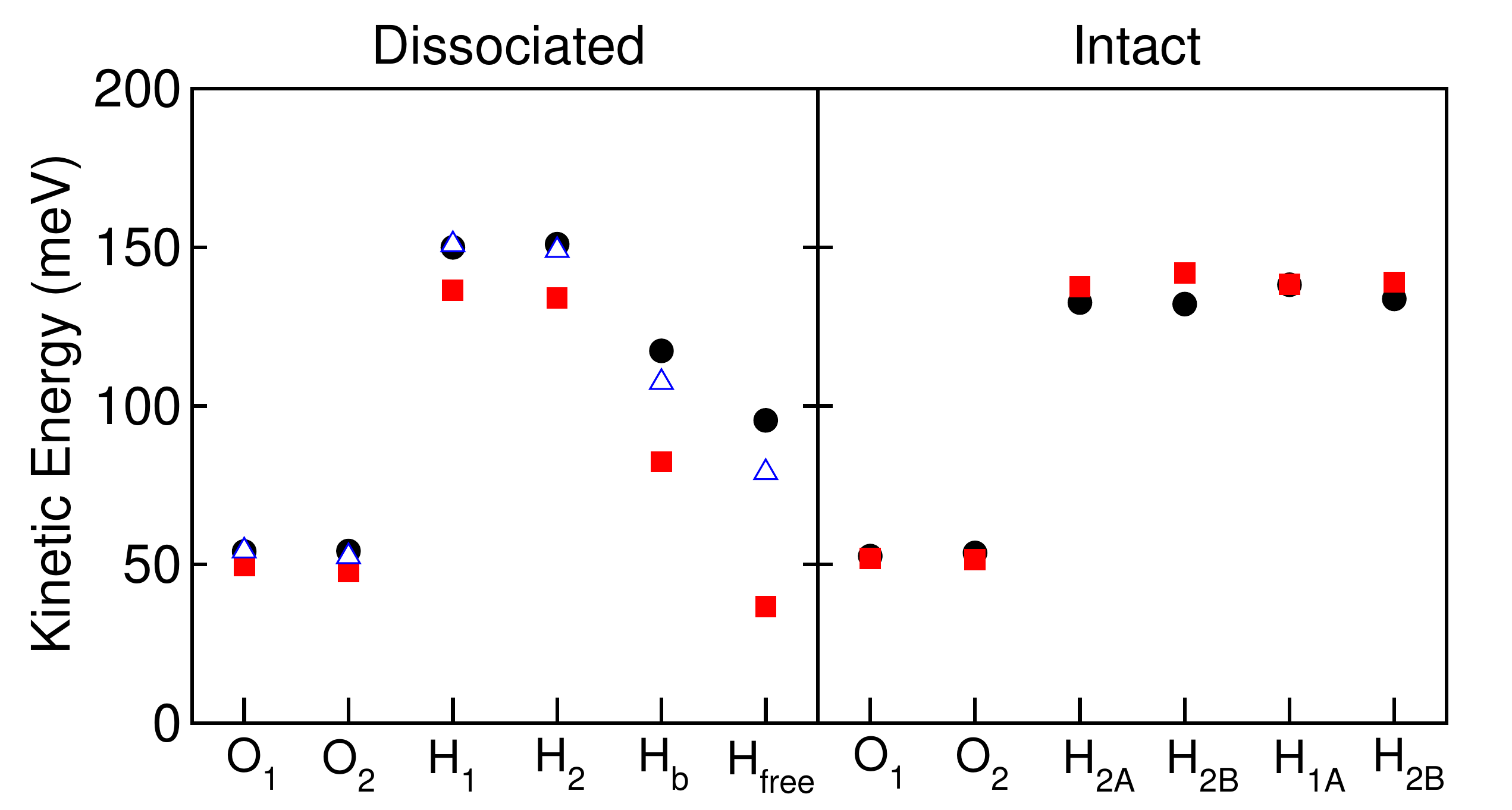}
    \caption{ Quantum kinetic energy calculated with the centroid virial estimator for the oxygen and hydrogen atoms in the dissociated (left) and intact (right) water wire simulations at 300~K. Black circles: no contraction, 6 beads, and PIGLET thermostat. Red squares: SLC to 1 bead (centroid contraction). Blue empty triangles: SLC to 4 beads.}\label{fig:kin_H}
\end{figure}

In Fig. \ref{fig:kin_H} we map these differences to the quantum kinetic energies of individual O and H atoms.
In the intact case the agreement between the non-contracted system and the centroid-contracted system relies minimally on an error cancellation: 
While the total quantum kinetic energy difference 
in the contracted and non-contracted system differs by only 3~meV (which is within our error bars), the difference projected only on the water molecules is 15 meV.
On the other hand, for the dissociated case, most of the difference in total quantum kinetic energy can be mapped to contributions of the H atoms, 
in particular the ones labeled H$_{\text{free}}$ and H$_B$ in Fig. \ref{fig:structures}.
This result can be traced back to the electronic structure and to the change in vibrational frequencies. 
In the dissociated case a large orbital hybridization takes place, so that the interfragment interaction is large and causes a large change in vibrational frequencies of the adsorbate with respect to the gas phase, thus making SLC unsuitable in this case. However, in the intact case, the interaction of the adsorbate and the surface is not so strong and the approximation works well. 
As discussed in the Appendix, contraction schemes are particularly problematic for cases in which the accuracy of the  potential used for the non-contracted part differs significantly between the states being compared. 
Therefore, in the following, we use our SLC scheme for the case where it works, namely the simulations with intact water molecules, and instead use only the PIGLET thermostat without the SLC scheme for the simulations with the dissociated species. 
It should also be stressed that even for the intact wire, the interaction with the surface is rather strong. We expect that the SLC should perform even better for weakly physisorbed systems.

We find that the quantum contribution to the kinetic energy, due to its local nature, is independent of the conformation of the water molecules. We used nine integration points, all calculated at 160K. 
In total, we have performed an equivalent of 1.1 ns of {\sl ab initio} simulations. We define the NQE contributions to the dissociation reaction as
\begin{equation}
\Delta \Delta F_d = \Delta F_q^{\text{diss}} - \Delta F_q^{\text{intact}}
\end{equation}
\noindent where $\Delta F_q$ is given by Eq. \ref{eq:TI} obtained from our PIMD simulations (fully anharmonic) and analytically from the harmonic approximation between the Izz and the Dt structures (by taking the difference between the free energy of coupled classical harmonic oscillators and quantum harmonic oscillators). Both curves have an integrand that is negative over the domain of integration (see SI), which implies that in both cases nuclear quantum effects stabilize the dissociated structure. 

We obtain 
\begin{eqnarray}
\Delta\Delta F_d^{\text{PIMD}} &=& -0.150 \pm 0.008 \, \text{eV} \nonumber\\
\Delta\Delta F_d^{\text{harm}} &=& -0.128 \, \text{eV}, \nonumber 
\end{eqnarray}
which represents an overall increase of about $20$ meV in the NQE contributions towards stabilization of the dissociated structures due to anharmonicity. 
In the harmonic case, where we can calculate the full quantum harmonic free energies of dissociation presented in Table \ref{tab:Dis_light}, we conclude that nuclear quantum effects are of the same magnitude as these differences themselves, and strongly stabilize dissociation. Taking the Izz$\to$Dt reaction as our example, the full (quantum) harmonic free energy of dissociation at 160~K is -70 meV and, as shown above, the quantum contributions to this free energy are -128 meV, meaning that the classical component of the free energy destabilizes dissociation by around 50 meV.
From the literature\cite{Sanchez:2010ew}, anharmonic (classical) free energies of dissociation of water adsorbed on surfaces are estimated to be $\sim$0.1 eV lower than the total energies of dissociation at the potential energy surface. These effects are of the same order of magnitude as our harmonic estimates in Table \ref{tab:Dis_light}. Thus, we can expect that free energy differences are of the same order of magnitude even in the 
anharmonic case (which is here challenging to calculate) and that, even though entropic effects due to temperature may favor the intact structures which have softer low-frequency modes for translations and rotations, NQE will still play a very important role in the energetic balance of these reactions. 

\subsection{Properties from Dynamics and Electronic Structure Interplay with NQE\label{sec:results:dyn} }

We now turn our attention to the interplay of NQE and the electronic structure, especially regarding the charge distribution on the adsorbates and changes in the work function of the surface.
For these calculations we have considered two different temperatures: 160 and 300 K. We note that for the higher temperature case we can sample conformational space faster and we have at least 10 ps of simulations (after thermalization) in all cases, 
while at the lower temperature we could only collect 5 ps of simulations, so that results should be considered only qualitative.

\begin{figure}[!hbt]
  \subfigure[160K]{
   \includegraphics[width=\columnwidth]{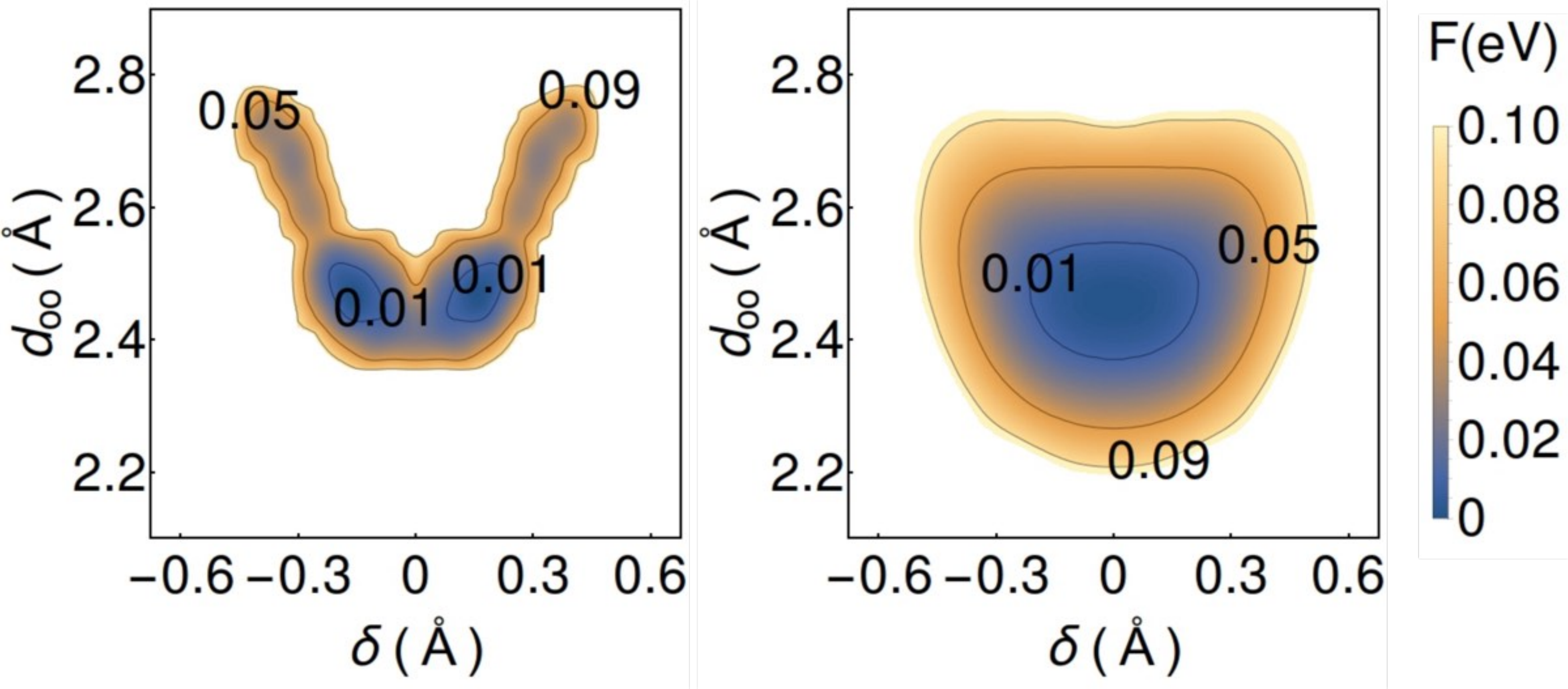}
   }
   \subfigure[300K]{
   \includegraphics[width=\columnwidth]{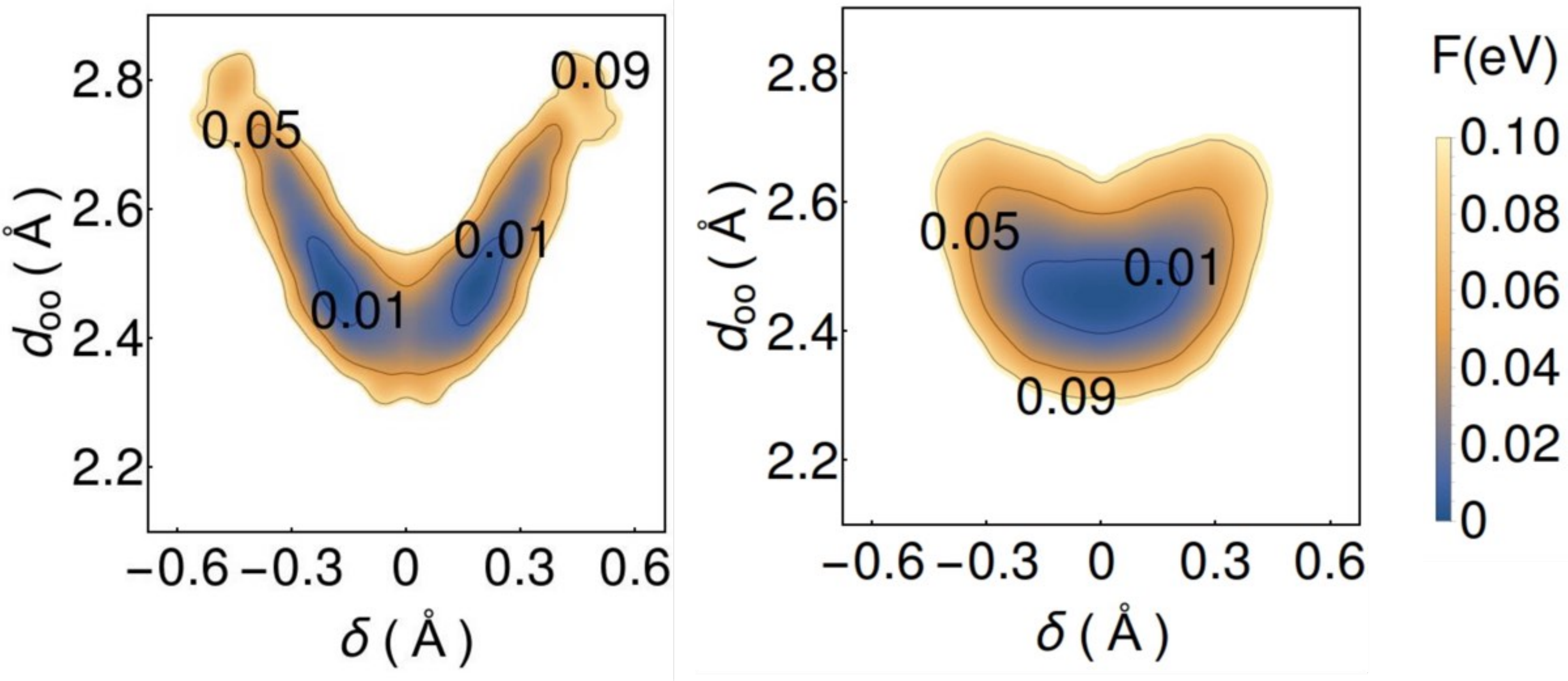}
   }
   \caption{Free energy profile of $d_{OO}$ vs $\delta$ for MD and PIMD simulations of the dissociated structures at 160K in (a) and at 300~K in (b).}   \label{fig:batman-symm}
\end{figure}

By comparing several different geometrical aspects of our MD and PIMD simulations, we find that the main qualitative difference
between the two cases happens (in agreement with other observations in the literature\cite{Michaelides_PRL_2010}) for the $\delta$ coordinate of the dissociated structures, 
where $\delta=\mathbf{d}_{OH}\cdot\mathbf{e}_{OO}-d_{OO}/2 $,  $\mathbf{e}_{OO}$ is the unit vector in the direction connecting two neighboring oxygen atoms and $\mathbf{d}_{OH}$ is one of the two O-H bond vectors.
In Fig. \ref{fig:batman-symm} we show a free energy profile of $d_{OO}$ vs. $\delta$ for MD and PIMD simulations at $T=160$ K and $T=300$ K.
The classical nuclei simulations
reveal a profile with two minima at both temperatures, indicating that the proton transfer is an activated process. When NQE are included, the barrier disappears, the proton becomes completely shared, and it is not anymore possible to distinguish H$_2$O from OH. 
We observe that the quantum distributions are dominated by ZPE at both temperatures, and the difference between classical and quantum nuclei follow the same trends at both temperatures as well.
In the following we show only results at 300~K due to the lack of statistical sampling in our 160K simulations.

\begin{figure}[!hbt]
\centering
\includegraphics[width=0.78\columnwidth]{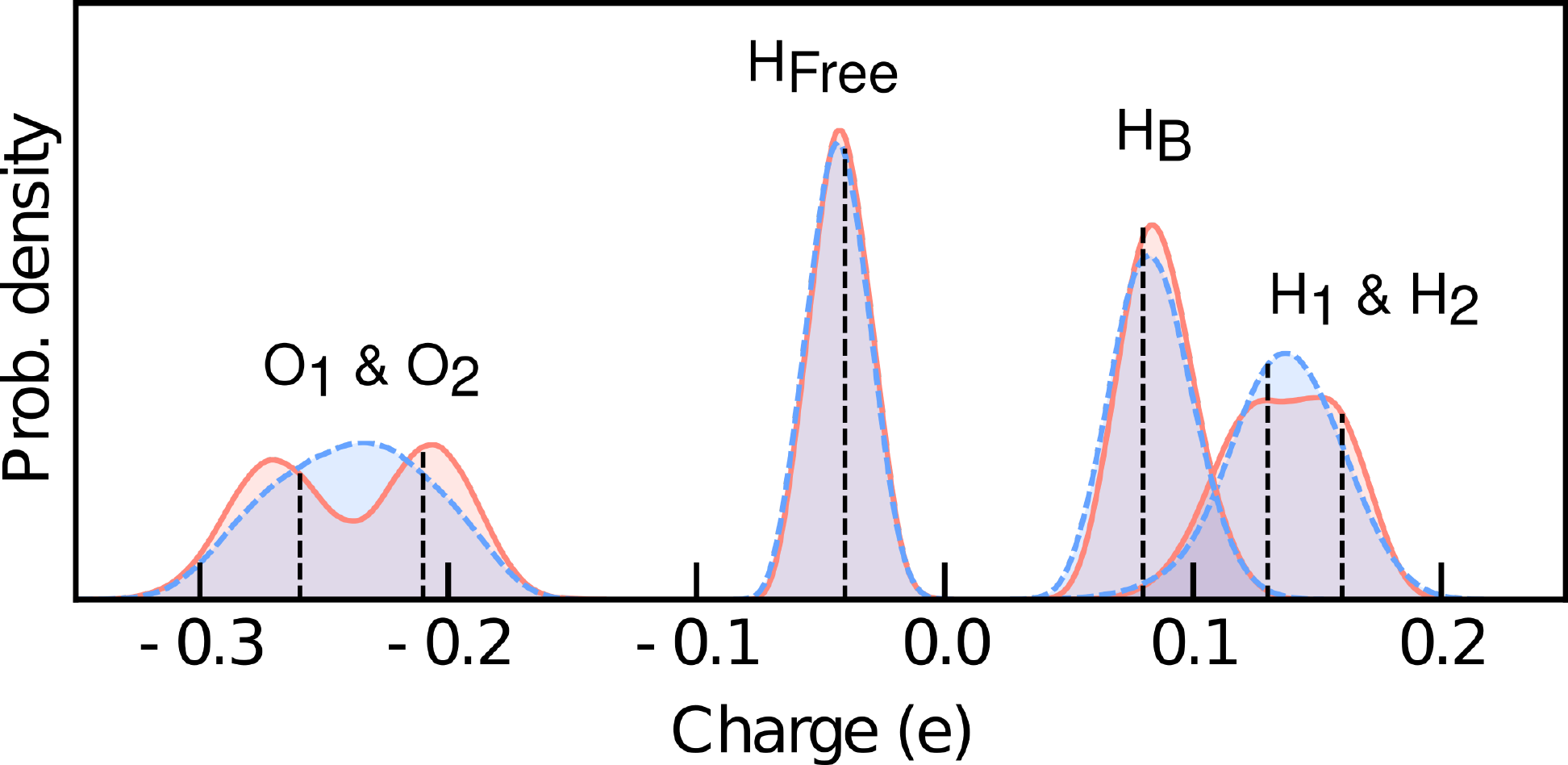}
    \caption{Probability density distribution of Hirshfeld charges for dissociated structures for MD simulations (red) 
    and PIMD simulations (blue ) at 300~K in. The vertical dashed lines show the charge value 
    for the minimum energy geometry.  Atoms labels are defined in Fig. \ref{fig:structures}.}\label{fig:dissoc_charges}
\end{figure}

\begin{figure}[!hbt]
   \centering
    \includegraphics[width=0.78\columnwidth]{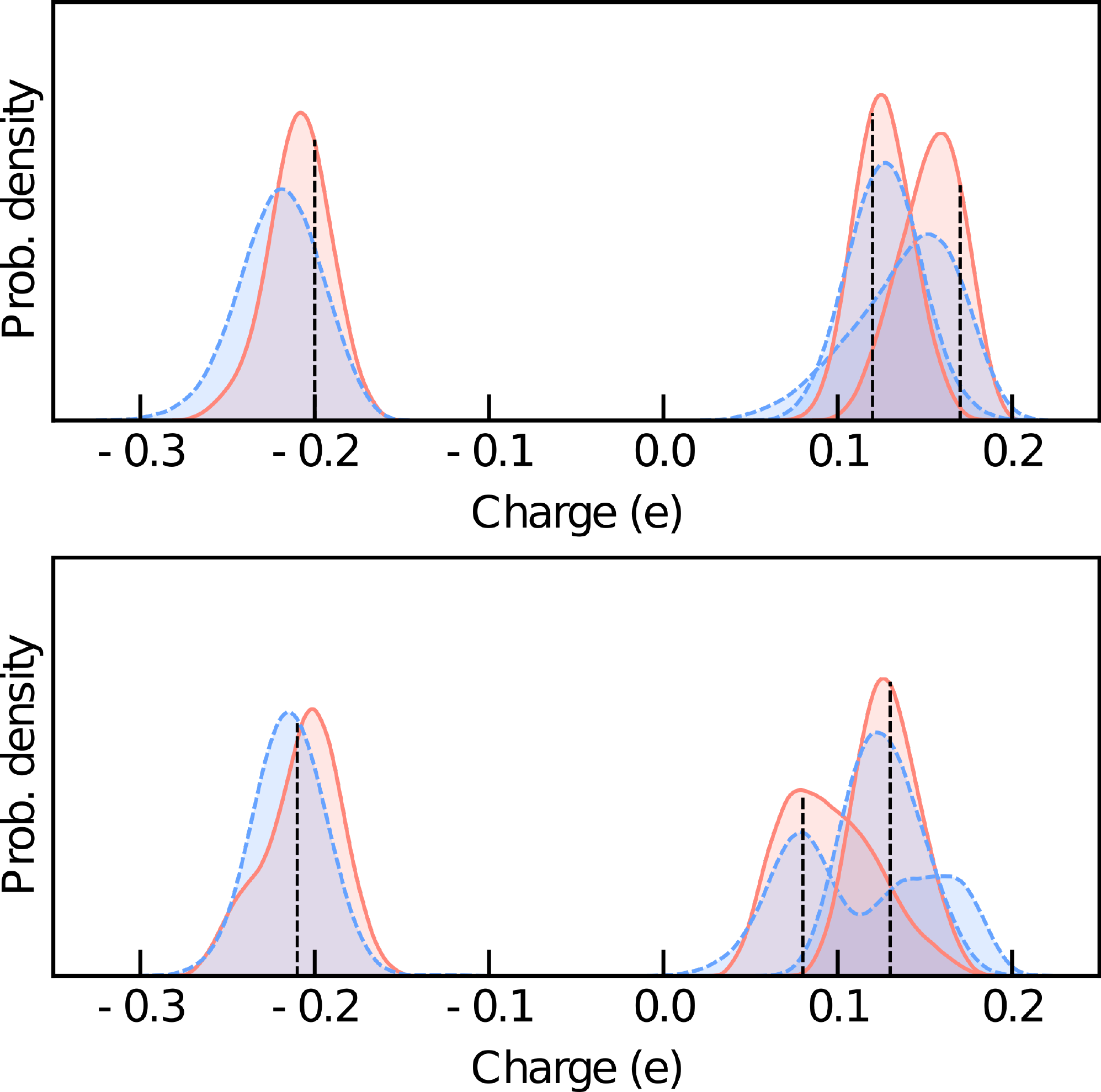}
    \caption{Probability density distribution of Hirshfeld charges for intact (Izz-like) structures for MD simulations (red) 
    and PIMD simulations (blue) at 300~K. The vertical dashed lines show the charge value for 
    the minimum energy geometry. Atoms labels are defined in  Fig. \ref{fig:structures}.} \label{fig:intact_charges}
\end{figure}

The Hirshfeld charge analyses for selected oxygen and hydrogen atoms in the dissociated and the intact structures are shown 
in Figures \ref{fig:dissoc_charges} and \ref{fig:intact_charges}, respectively.
We first analyze the dissociated case, shown in Fig. \ref{fig:dissoc_charges}. In agreement with the geometrical analysis, the charge on the oxygen atoms
shows a similar behaviour as the asymmetric $\delta$ coordinate, i.e. two peaks in the MD simulations and only one for PIMD case. 
This gives additional support for the complete delocalization of the shared hydrogen forming a H$_3$O$_2$-like complex. Interestingly, the dissociated hydrogen does not show an appreciable difference in the charge state 
when including NQE.
In the intact case (Fig. \ref{fig:intact_charges}), the main differences emerge on the hydrogens which are not involved in a hydrogen bond (H$_{1A}$ and H$_{2A}$). 
The inclusion of NQE allows them to explore an extended configurational space where the interaction with the non-equivalent sites in the vicinity of the step produce a charge fluctuation.
In particular $H_{1A}$ presents a long tail to lower value of charge and $H_{2A}$ shows a bimodal distribution,
where the higher charge peak corresponds to the situation where that H points perpendicular to the terrace. We also note that most of the average charge distributions are displaced from the value calculated at the minimum energy structure (also shown as vertical lines in the figures) -- an effect that has also been observed in completely different systems \cite{Schran_CPL_2017}. This further underlines the importance of explicit inclusion of entropy at finite temperatures when studying electronic properties of flexible, H-bonded systems. 

\begin{figure}[!hbt]
    \includegraphics[width=\columnwidth]{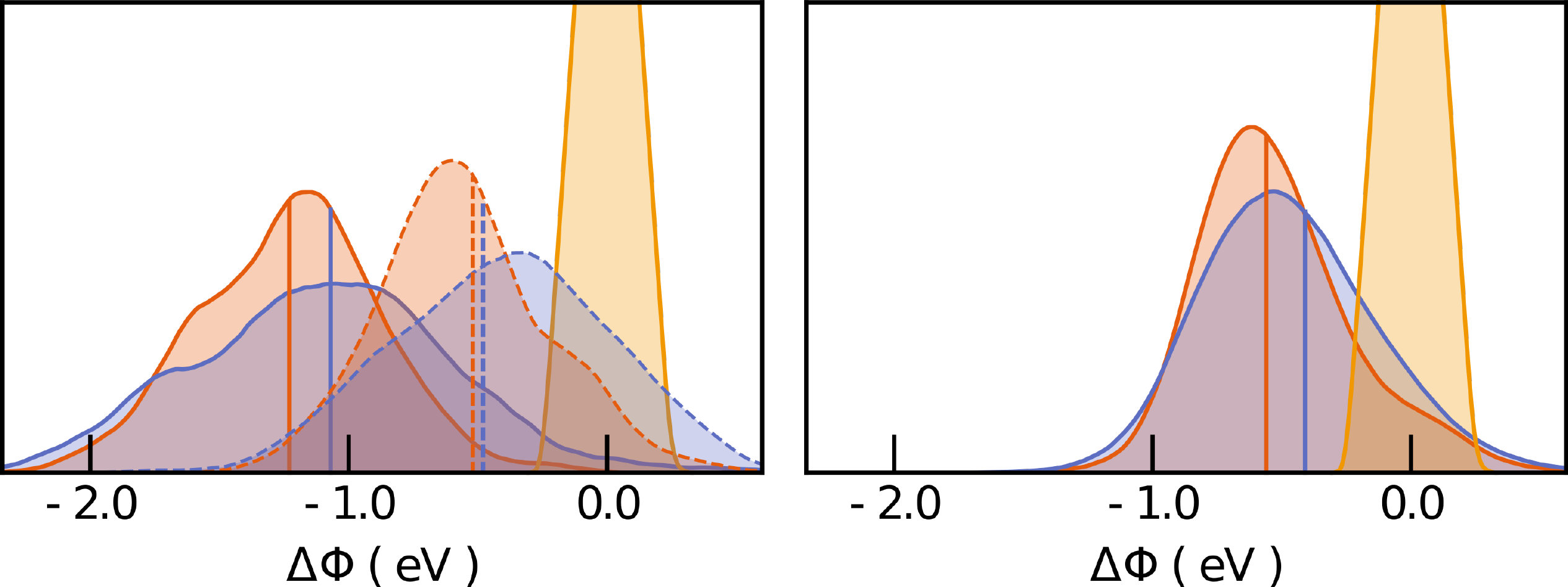}
    \caption{Probability density distribution of work function changes as obtained from MD simulations (red) and PIMD simulations (blue) at 300~K. The yellow curve correspond to MD simulations of Pt surface without water molecules.
    The left plot corresponds to the intact, Izz-like (solid lines) and Ilw-like (dashed lines) structures and the right plot to the dissociated structures. The vertical lines show the mean values of the work function for each case.} \label{fig:work}
\end{figure}

Finally, we analyze changes in the work function of our supercell for 
the different states of the adsorbate, and also when
including temperature and nuclear quantum effects. We show in Table \ref{tab:work} the calculated work function changes $\Delta\Phi$ when the adsorbates are added to the surface, and in Fig.~\ref{fig:work} the distribution of $\Delta\Phi$ collected from MD and PIMD simulations. The adsorption causes, on 
average, a decrease
of the work function in both intact and dissociated cases, which is very accentuated for  
the Izz structures. Comparing the distributions shown in Fig.~\ref{fig:work} obtained from
simulations with classical nuclei and quantum nuclei, we observe that NQE 
cause a broadening and a shift of the distributions to less negative values. Compared to the relaxed geometries, the inclusion of temperature (classical nuclei) shifts the average value of $\Delta\Phi$ by about 0.2 eV for the intact structures, and 0.1 eV for the dissociated ones. Inclusion of NQE further shifts this value by 0.15 eV for intact structures and only 0.05 eV for the dissociated ones. This means that full inclusion of temperature and NQE can decrease work function changes by up to 0.4 eV. 
The magnitude of the observed changes in the work function value may not 
be a common feature of all surfaces, since we expect stepped surfaces to be particularly affected due to the strong surface dipole generated at the step by the Smoluchowski 
effect\cite{Smolu_PR_1941}. Since work functions can be measured locally on steps\cite{JiaSakurai1998}, the effect we report is something that could be measured.

\begin{table}[ht]
\caption{\label{tab:work} Work function changes, in eV, when different structures are added to the Pt surface. Values reported from the MD and PIMD simulations are averages through our simulations at 300~K.}
\begin{tabular}{cccc}
\hline
Structure & $\Delta \Phi^{\text{static}}$ & $\Delta \Phi^{\text{MD}}$ & $\Delta \Phi^{\text{PIMD}}$ \\
\hline
\hline
Izz & -1.51	&	-1.23	&	-1.07 \\
Ilw & -0.72	&	-0.56	&	-0.41 \\
Dh  & -0.62	&			&         \\
Dt  & -0.61	&	-0.52	&	-0.48 \\
\hline
\end{tabular}
\end{table}

\begin{figure}[!hbt]

    \includegraphics[width=\columnwidth]{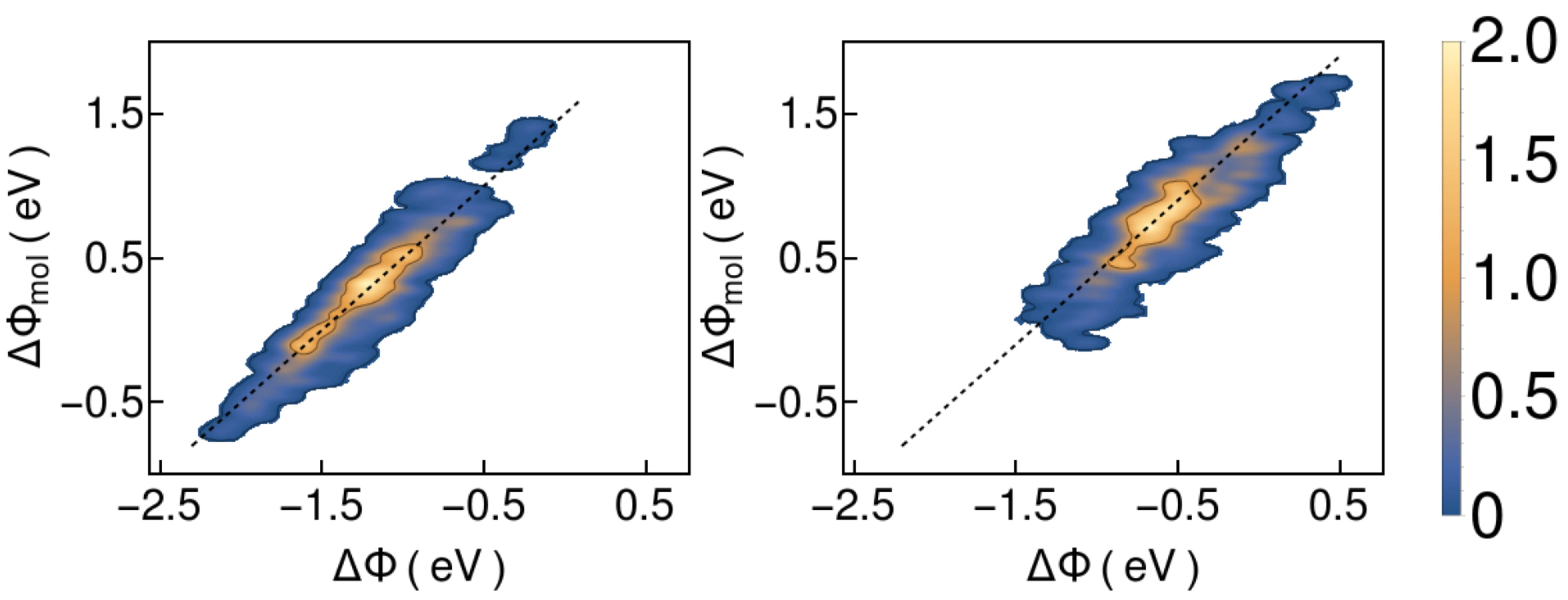}
   \caption{Probability density distribution of the work function change due to molecular dipole ($\Delta\Phi_\text{mol}$)  vs $\Delta\Phi$ for MD simulations of the intact structures 300~K. Left: Izz-like structures. Right: Ilw-like structures. The dotted lines are a guide to the eye with a positive slope of 1. } \label{fig:dipole}
\end{figure}

In order to explain the origin of the large $\Delta\Phi$ we observe, we show in Fig. \ref{fig:dipole} the correlation between $\Delta\Phi$ and the work function change caused by the molecular dipole perpendicular to the surface $\Delta\Phi_{\text{mol}}$. Following references \cite{LeungChan2003, HeimelBredas2007, Hofmann_NanoLett_2010, VerwusterZojer2015} one can model the total work function change as $\Delta\Phi = \Delta\Phi_{\text{mol}} + \Delta\Phi_{\text{x}}$, where $\Delta\Phi_{\text{mol}}=-e \mu_{mol} / A \epsilon_0$ ($A$ is the area of the surface in the unit cell and $\mu_{mol}$ is the molecular dipole perpendicular to the surface), and  $\Delta\Phi_{\text{x}}$ correspond to other contributions to the work function change
We calculate $\Delta\Phi_{\text{mol}}$ with a procedure detailed in the SI. 
The correlation we observe in Fig. \ref{fig:dipole} is very close to linear and with a slope of 1, meaning that other contributions to work function changes ($\Delta\Phi_{\text{x}}$) act just as an additive constant, which we find to correlate with the average charge transferred to (or from) the adsorbated molecules (see Fig. S4). 

Given the quantitative analysis of the work function in the presence of different structures of the wire, we propose that the experimental determination of the local change in work function could be used to discriminate between different configurations. 
For the Izz structure we predict a reduction in the work function of about 1 eV. Both the Ilw and the dissociated structure should yield a reduction of about 0.5~eV. 
Deuteration experiments could discriminate between these two scenarios, since NQE produce almost no work function change for the dissociated structures, but an appreciable one for the intact structures.

\section{Conclusions \label{sec:conclusions}}

In this paper we have studied the impact of nuclear quantum effects in many different aspects of water dissociation  
on the stepped Pt(221) surface at finite temperatures. Our study using DFT shows that this system can explore several different
local minima, and due to the low barriers connecting these minima, anharmonicities and nuclear quantum effects (NQE) are likely to be important. In particular for intact water molecules we find structures in which the oxygens of the water molecules are
not aligned, forming a zigzag-like H-bonded structure, and also structures in which the oxygens are aligned, forming a linear H-bonded structure. When water dissociates the dissociated hydrogen atom can occupy both hollow and top positions, but actually prefers the top position close to the step.
We have shown that NQE make all these conformations thermodynamically competitive, favoring in particular partial dissociation of water. Indeed through the calculation of 
reaction paths and rates of dissociation (still in the harmonic approximation, but for different conformations), we have shown that the inclusion
of ZPE can increase the dissociation rate by three orders of magnitude. We estimated the crossover temperature for these reactions to be around 130K,
so that ZPE, and not tunneling, should be the most important NQE at all temperatures around or above this one. Regarding the inclusion of van der Waals
corrections, we observe that it can have a non-negligible effect in stabilizing intact structures, since it makes hydrogen bonds stronger. 
Perhaps more importantly, we observe that the inclusion of vdW lowers the dissociation barrier 
by 60 meV, which has a truly non-negligible impact in dissociation rates.

In order to perform {\sl ab initio} simulations including NQE including anharmonicities and seamlessly taking into account the relative stability
of different low energy structures at a lower cost,
we proposed a simple scheme that is applicable when a natural partition of the system is possible. This is precisely the situation when
molecules are weakly bound on surfaces. In our scheme, we perform a ring polymer contraction in the full system and calculate only the adsorbate
with the full path integral treatment but in isolation. We showed that this procedure works very well when one is interested in studying NQE on
a weakly bound adsorbate, but that when strong orbital hybridisation and thus chemisorption occurs (changing dramatically the higher vibrational frequencies
of the adsorbate with respect to the gas phase situation), this scheme introduces a large error. We explained how this error arises in general
ring-polymer contraction schemes by making considerations on model harmonic systems. Then, using PIMD simulations we performed quantum to classical
thermodynamic integrations and found that the NQE contributions to the dissociation free energies amount to 150 meV per dimer, favoring
the dissociated state. One can understand most of this effect by performing an analysis of the change in vibrational frequencies of the adsorbate
before and after dissociation: A red shift of more than 1000 cm$^{-1}$ of the vibrational frequency from the OH to the Pt-H stretching vibration, as well as a red-shift
of all other peaks due to the formation of stronger non-covalent bonds upon dissociation explain the lowering of ZPE.

Finally, we analyzed the interplay between NQE and the electronic structure. In particular, we find that
work function changes on these steps are significantly impacted by both temperature and NQE, with changes up to 0.4~eV. We propose that our quantitative assessment of the local change in work function, together with the observation of the stronger NQE for the intact than for the dissociated structures, could provide an approach to experimentally verify the structure of water wires on Pt(221) and probably other metallic stepped surfaces.

\section{Acknowledgements}

M.R. and Y.L. thank Luca Ghiringelli for useful discussions. M.C. acknowledges funding from the Swiss National Science Foundation (project ID 200021-159896).
The authors acknowledge computer time from the Swiss National Supercomputing Centre (CSCS), under project number s719 and s711.

\appendix

\section{An estimate of errors in ring-polymer contraction schemes} \label{app:rpc-harmonic}
In order to understand the impact of the SLC approximation on the accuracy of the
estimates of quantum mechanical observables it is useful to consider a simple
model, in which the presence of the substrate determines a shift in the harmonic
frequency of the normal modes of the molecule. 
In other terms, we consider $V_\tm(q)=\frac{1}{2}m\omega_\tm^2 q^2$ to be the gas-phase potential, and $V_\tf(q) = \frac{1}{2} m\omega_\tf^2 q^2$ to be the full potential describing the vibrations of the molecules in contact with the surface.
It's worth stressing that while we use a notation that reflects this specific 
application, the following considerations apply equally well
to the general case of RPC~\cite{mark-mano08jcp}, with a reference potential that has a frequency
$\omega_\tm^2$ and is used to approximate an accurate potential with frequency $\omega_\tf^2$.
Since both harmonic potentials are diagonal in the ring-polymer normal mode basis $\left\{\tilde{q}^{(k)}\right\}$, 
one can rewrite the overall potential energy term as 
\begin{equation}
V_P(\bq) = \sum_{k=1}^{P} \frac{1}{2}m\omega_\tm^2 \left[\tilde{q}^{(k)}\right]^2 
+ \sum_{k=1}^{P'} \frac{1}{2}m (\omega_\tf^2 - \omega_\tm^2) \left[\tilde{q}^{(k)}\right]^2.
\end{equation}
One sees that the error relative to the case in which the whole system is treated
using $V_\tf$ is restricted to the normal modes between $P'+1$ and $P$. 
Given that the expectation value of of the potential in the molecule is given by $m\omega^2_\tm\left< \left[\tilde{q}^{(k)}\right]^2 \right>_\tm = P k_B T/\left<{\omega_k^2 + \omega_\tm^2}\right>$,
and assuming that $\left|\omega_\tf^2-\omega_\tm^2\right|\ll \omega_\tf^2+\omega_k^2$, 
one sees that the error amounts to
\begin{equation}
\epsilon_\text{SLC}=\left(1-\frac{\omega_\tm^2}{\omega_\tf^2}\right) \frac{k_B T}{2} \sum_{k=P'+1}^P \frac{\omega_\tf^2}{\omega_k^2 + \omega_\tf^2}.
\end{equation}
Thus, the error is a fraction of the residual resulting from a simulation of 
the adsorbed molecule with $P'$ beads and no contraction (that we shall call
$\epsilon_{P'}$ in what follows) that reduces to zero when the adsorption
process does not modify a given vibrational mode and it is more important for higher $\omega_\tf$ . 
If one computes the full potential (FP) on top of RPC sampling (a common strategy
to reduce the error in a RPC simulation\cite{mark-mano08cpl}), the error term involves 
the next-order term in $\omega_\tf^2-\omega_\tm^2$:
\begin{equation}
\epsilon_\text{SLC,FP}=\left(1-\frac{\omega_\tm^2}{\omega_\tf^2}\right) \frac{k_B T}{2} \sum_{k=P'+1}^P \left[\frac{\omega_\tf^2}{\omega_k^2 + \omega_\tf^2}\right]^2.
\end{equation}

A final, important observation concerns the use of SLC (and RPC in general) to
compute the differences in quantum properties between two different states. 
Using a tilde symbol for all quantities that refer to this second state
(e.g. a different surface, or a different conformer of the molecule), the error
when performing a SLC calculation is
\begin{equation}
(\epsilon_{P'}-\tilde{\epsilon}_{P'}) \left(1-\frac{\omega_\tm^2}{\omega_\tf^2}\right) +
\tilde{\epsilon}_{P'}\left(\frac{\tilde{\omega}_\tm^2}{\tilde{\omega}_\tf^2}-\frac{\omega_\tm^2}{\omega_\tf^2}\right).
\end{equation}

The second term means that if the reference potential is not equally good for the two
states, the potential-energy difference could have a larger error than if one had done 
a  low-$P$ simulation with the full potential, which could have benefitted from error cancellation.

\bibliographystyle{aipnum4-1}

\end{document}